\input lecproc.cmm
\contribution{Non--Linear Evolution of Cosmological Perturbations}
\author{Sabino Matarrese}
\address{Dipartimento di Fisica `Galileo Galilei', Universit\`a di 
Padova \newline
via Marzolo 8, I--35131, Padova, Italy}

\abstract{In these lecture notes I review the theory of the non--linear 
evolution of cosmological perturbations in a self--gravitating collisionless 
medium, with vanishing vorticity. The problem is first analyzed in the 
context of the Newtonian approximation, where the basic properties of the 
Zel'dovich, frozen--flow and adhesion algorithms are introduced. An exact 
general relativistic formalism is then presented and it is shown how the 
Newtonian limit, both in Lagrangian and Eulerian coordinates, can be
recovered. A general discussion on the possible role of possible relativistic 
effects in the cosmological structure formation context is finally given.}

\titlea{1}{Introduction}

An important theoretical issue in cosmology is to understand the physical 
processes that occurred during the gravitational collapse of the matter which 
gave rise to the observed large--scale structure of the universe. A 
complementary issue is to reconstruct the initial conditions of the clustering 
process, e.g. the value of the cosmological parameters, the type of dark 
matter, the statistics of the primordial perturbations, starting from 
observational data such as the spatial distribution of galaxies or their 
peculiar velocities. Much work has been recently done in the latter direction, 
since more and more data on peculiar velocities of galaxies, as well 
as very large and complete galaxy redshift surveys have become available. 

A widely applied and well--motivated approximation when dealing with the 
dynamics of dark matter, either cold or hot, is to treat it as a system of 
particles having negligible non--gravitational interactions, a
self--gravitating collisionless system. 
The dynamics of such a system is usually approached by different 
techniques, depending on the specific application. For instance, the evolution 
of small perturbations on a Friedmann--Robertson--Walker (FRW) 
background is followed by analytical methods. 
The non--linear evolution in cases where some symmetries are present
can also sometimes be followed analytically: typical examples being the 
spherical top--hat model in the frame of the Newtonian approximation 
and the 
Tolman--Bondi solution in General Relativity (GR). 
There are also useful approximations valid in the 
mildly non--linear regime, such as the Zel'dovich 
approximation (Zel'dovich 1970). Besides this classical approach, a number 
of variants have been proposed, all trying to overcome the
inability to follow the development of 
structures beyond caustic formation. 
Different approximations apply in the highly non--linear regime, such as the 
hierarchical closure ansatz for the BBGKY equations (e.g. Peebles 1980) . 
The most general problem of studying the fully non--linear dynamics 
of a collisionless system in Newtonian theory can only be 
followed by numerical techniques, such as $N$--body codes (e.g. Hockney \&
Eastwood 1981; Efstathiou et al. 1985). 
Finally, the non--linear relativistic evolution of 
a pressureless fluid has been recently studied in a number of papers (e.g. 
Matarrese \& Terranova 1995, and references therein). 

Here we review various methods, trying to show their possible 
interconnections. Section 2 presents 
the Zel'dovich, frozen--flow and adhesion approximations. 
Section 3 deals with the relativistic dynamics of 
a self--gravitating collisionless and irrotational fluid within GR. 
A final general discussion is given in Section 4.

\titlea{2}{Newtonian dynamics of self--gravitating collisionless matter}

\titleb {2.1}{General formalism}

Let us start by writing the Newtonian equations for the evolution of 
collisionless particles in the expanding universe (e.g. Peebles 1980). 
These can be written using suitably rescaled variables and in 
comoving coordinates. We shall assume that the universe is spatially flat 
and matter dominated, so that the scale factor reads 
$a(t)=a_0(t/t_0)^{2/3}$ (a subscript $0$ will be used to define quantities at 
some ``initial time" $t_0$). A generalization of these formulae to the open 
(and closed) universe case is given in (Catelan et al. 1995). 
The Euler equations read
$$
{d {\bf u}\over d a} + {3 \over 2 a} {\bf u} = - {3 \over 2 a} 
\nabla \varphi, 
\eqno(1)
$$
where ${\bf u} \equiv d {\bf x}/ d a$ 
is a rescaled comoving peculiar 
velocity field. The symbol ${d \over d a}$ stands for the total (convective) 
derivative 
$$
{d \over d a}= {\partial \over \partial a} + {\bf u} \cdot
\nabla.
\eqno(2)
$$

The continuity equation can be written in terms of
the comoving matter density $\eta({\bf x},t) \equiv \varrho({\bf x}, t)
~a^3(t) / \varrho_0 a_0^3$ (where $\varrho_0$ is the mean mass 
density at $t_0$)
$$
{d \eta \over d a} + \eta \nabla \cdot {\bf u} = 0, 
\eqno(3)
$$
while the rescaled local (or peculiar) gravitational potential 
$\varphi \equiv (3t_0^2/2a_0^3) \varphi_g({\bf x},t)$ is 
determined by local density inhomogeneities $\delta({\bf x},t) \equiv 
\eta({\bf x},t) - 1$ through the Poisson equation
$$
\nabla^2 \varphi = {\delta \over a}.
\eqno(4)
$$
We can restrict the analysis to 
initially irrotational flow. According to Kelvin's circulation theorem, 
in the absence of dissipation, vorticity is conserved along each fluid 
trajectory; in particular, a fluid with vanishing initial vorticity will 
forever remain irrotational. However, for a collisionless fluid such a property 
breaks down after caustic formation: a vorticity component is 
created in multi--stream regions, simply because the local Eulerian 
velocity field takes contributions from different Lagrangian fluid elements 
at the same position. 
Outside the regions of orbit mixing and/or after applying a suitable low--pass 
filter one can define a velocity potential by ${\bf u}({\bf x},a) = 
\nabla \Phi({\bf x},a)$. It is then easy to derive from the Euler 
equation a Bernoulli equation relating the velocity potential $\Phi$ to 
the gravitational one $\varphi$, 
$$
{\partial \Phi \over \partial a} + {1 \over 2} 
\bigl( \nabla \Phi \bigr)^2 = - {3 \over 2 a} (\Phi + \varphi).
\eqno(5)
$$

By integrating along the particle trajectory ${\bf x}(a)$ 
one finds a formal solution of the continuity equation,
$$
\eta({\bf x}, a) = \eta_0({\bf q}) \exp{\biggl\{ - \int_{a_0}^a da' 
\nabla \cdot {\bf u}[{\bf x}({\bf q},a'),a'] \biggr\}}.
\eqno(6)
$$
This can be compared with the formula obtained from mass
conservation, $\eta({\bf x},a) d^3 x = \eta_0({\bf q}) d^3 q$, 
where ${\bf q}$ is the initial (Lagrangian) position of the particle which has 
reached the (Eulerian) position ${\bf x}$ by the time $a(t)$.
One has either the well--known Lagrangian form 
$$
\eta[{\bf q}({\bf x}, a),a] = \eta_0({\bf q}) |\!| 
\partial x / \partial q |\!|^{-1},
\eqno(7)
$$
where $|\!|\partial x / \partial y|\!|$ is the Jacobian 
determinant of the transformation $x \to y$, or the Eulerian one
$$
\eta({\bf x}, a) = \eta_0({\bf q}) |\!| \partial q / \partial x |\!|,
\eqno(8)
$$
which however requires inverting the trajectory to find
${\bf q}({\bf x},a)$.
These solutions are only valid as long as 
no shell--crossing (caustic) has occurred, so that there is a one--to--one 
correspondence between Lagrangian and Eulerian positions. 
Before caustic formation all these forms are identical. 

\titleb {2.2}{Zel'dovich approximation}

The simplest approximation is of course the linear one, which 
consists in neglecting the terms
${\bf u} \cdot \nabla {\bf u}$ in the Euler equations and 
$\nabla \cdot (\delta {\bf u})$ in the continuity equation.
The resulting expressions read 
${\bf u}_{LIN}({\bf x},a) = - \nabla \varphi_0({\bf x})$, 
$\varphi_{LIN}({\bf x}, a) = \varphi_0({\bf x})$ and
$\eta_{LIN}({\bf x},a) = 1 + a \nabla^2 \varphi_0({\bf x})$, 
having neglected the contribution of decaying modes.

The next step is the Zel'dovich approximation (ZEL), based 
on the ansatz of extrapolating the equation ${\bf u} = - \nabla 
\varphi$ (i.e. $\Phi = - \varphi$) beyond linear theory; replacing this ansatz 
into the Euler equations gives
$$
{d {\bf u} \over d a} = 0, 
\eqno(9)
$$
which has to be solved together with Eq.(3).
The resulting system can be taken as the definition of ZEL.
In this approximation a particle initially placed in ${\bf q}$ moves
along a straight--line with constant ``speed" ${\bf u}$ 
determined by the value of the initial peculiar gravitational potential 
in ${\bf q}$, namely
$$
{\bf x}({\bf q}, \tau_a) = {\bf q} - \tau_a \nabla_{\bf q} 
\varphi_0({\bf q}), 
\eqno(10)
$$
with $\tau_a \equiv a - a_0$. The 
velocity field is conserved along each particle trajectory:
$$
{\bf u}({\bf x},\tau_a) = {\bf u}_0({\bf q}) = - 
\nabla_{\bf q} \varphi_0({\bf q}).
\eqno(11)
$$
The velocity potential, defined by 
${\bf u}({\bf x},\tau_a) =  \nabla_{\bf x} 
\Phi({\bf x},\tau_a) = \nabla_{\bf q} \Phi_0({\bf q})$, which 
obeys the Hamilton--Jacobi equation 
$$
{\partial \Phi \over \partial \tau_a} + {1 \over 2} 
\bigl( \nabla_{\bf x}
\Phi\bigr)^2 = 0.
\eqno(12)
$$
The solution of Eq.(12) is
$$
\Phi({\bf x},\tau_a) = \Phi_0({\bf q}) + ({\bf x} - {\bf q})^2 
/2 \tau_a.
\eqno(13)
$$
The density field is usually represented in the Lagrangian form 
$\eta({\bf q},\tau_a) = \eta_0({\bf q}) |\!|
{\bf 1} + \tau_a {\bf D}_0({\bf q}) |\!|^{-1}$,
where ${\bf 1}$ is the unit matrix and ${\bf D}_0$ the 
deformation tensor, with components $D_{0,ij}({\bf q}) = 
\partial^2 \Phi_0({\bf q}) /\partial q_i \partial q_j$.
The deformation tensor can be locally diagonalized, by going to principal 
axes $X_1,X_2,X_3$, with eigenvalues $\alpha_1,\alpha_2,\alpha_3$. We thus have
$$
\eta({\bf q},\tau_a) = { \eta_0({\bf q}) \over
\bigl(1 + \tau_a \alpha_1({\bf q})\bigr)
\bigl(1 + \tau_a \alpha_2({\bf q})\bigr)
\bigl(1 + \tau_a \alpha_3({\bf q})\bigr) }.
\eqno(14)
$$
According to the latter expression a singularity (caustic) in Lagrangian space 
would form at every point ${\bf q}$ where at least one eigenvalue,
say $\alpha_1$, is negative. 

Besides being, by construction, consistent with the growing mode 
of linear perturbations at early times,
ZEL provides a good approximation 
up to the time of first shell crossing.
The inconsistency of ZEL can be seen as follows.
Inserting the ansatz $\Phi=-\varphi$ into the Poisson equation one gets an
expression for the density fluctuation, $\delta_{DYN} = - a 
\nabla \cdot {\bf u}$, which is nothing but the linear 
theory relation between peculiar velocity and density fluctuation.
This point has been discussed by Nusser et al. (1991), who
refer to this determination of the density, $\eta_{DYN}=1+\delta_{DYN}$, as 
dynamical density, to distinguish it from the continuity density 
obtained from Eq.(14). It is possible to go further this way by replacing this
expression into Eq.(3): one gets the mass density in Lagrangian form 
$\eta_{DYN}({\bf q},\tau_a) = \eta_0({\bf q}) /(1 - \tau_a \delta_+({\bf q}))$,
where $\delta_+({\bf q}) = -(\alpha_1({\bf q}) + \alpha_2({\bf q}) + 
\alpha_3({\bf q}))$, 
and the Poisson equation was used to relate $\varphi_0$ to $\delta_0$ and 
we defined the (scaled) initial growing mode $\delta_+\equiv \delta_0/a_0$. 

It is then clear that the Zel'dovich ansatz is only exact 
for one--dimensional perturbations, where the two above 
expressions for the density coincide; in the general three--dimensional case it 
fails. 
An alternative understanding of the Zel'dovich approximation might be obtained 
in the frame of perturbation theory in Lagrangian coordinates. This approach 
has been recently reviewed by Buchert (1995). 

\titleb {2.3}{Frozen--flow approximation}

The frozen--flow approximation (FFA) (Matarrese et al. 1992) can be 
defined as the 
exact solution of the linearized Euler equations, where in the r.h.s. the 
growing mode of the linear gravitational potential is assumed.
Such an equation is solved by 
${\bf u}({\bf x},\tau_a) = {\bf u}_0({\bf x}) = - \nabla_{\bf x} 
\varphi_0({\bf x})$, plus a negligible decaying mode. 
In this approximation the peculiar velocity field ${\bf u}({\bf x},a)$ is
frozen at each point to its initial value, that is 
$$
{\partial {\bf u} \over \partial \tau_a} =0,
\eqno(15)
$$
which is the condition for steady flow. 
Such an equation can be used, together with the continuity 
equation to define FFA.
Particle trajectories in FFA are described by the integral 
equation
$$
{\bf x}({\bf q},\tau_a) = {\bf q} - \int_0^{\tau_a} d \tau_a' \nabla_{\bf x}
\varphi_0[{\bf x}({\bf q},\tau_a')] :
\eqno(16)
$$
particles during their motion update at each infinitesimal
step their velocity to the local value of the linear velocity field, without 
memory of their previous motion, i.e. without inertia. 
This would be the case of a particle moving under the influence of
a force in a medium with very large viscosity:
the damping here is determined by the 
Hubble drag while the force is the gravitational one. 

As we shall see shortly, no caustics are formed at finite time in FFA, 
so that all expressions for the density can be used interchangeably. 
It is nevertheless interesting to write the comoving mass density 
as given by Eq.(6), namely
$$
\eta({\bf x}, \tau_a) = \exp \int_0^{\tau_a}
d\tau_a' \delta_+[{\bf x}({\bf q},\tau_a')] ,
\eqno(17)
$$
having assumed $\eta_0({\bf x}_\star) \approx 1$. 
The logarithm of the density in ${\bf x}$
is given by the integral of the linear density field over the trajectory
of the particle which has arrived to ${\bf x}$ at time $\tau_a$, starting from
the Lagrangian position ${\bf q}$. 

FFA is, by construction, consistent with linear theory, so it follows
correctly the evolution at early times (precisely as it happens for ZEL).
The assumption of keeping the linear approximation for the 
velocity potential beyond the linear regime is justified by the
fact that this quantity is more sensitive to large wavelength modes
than the density, it is therefore less affected by strongly non--linear 
evolution.
Stream--lines are frozen to their initial shape, so multi--stream regions
cannot form, unless they were already present in the initial velocity field. 
FFA therefore avoids the formation of caustics at finite $\tau_a$, so one can 
try to extrapolate the approximation after the time 
at which the first shell--crossing would have appeared 
according to ZEL. A particle moving according to FFA
has zero component of the velocity in a place where the same
component of the initial gravitational force is zero, it will then
slow down its motion in that direction approaching 
such a position: particles in FFA need infinite time to reach those 
places where a pancake, a filament or a knot will occur.
Moreover, since, unlike ZEL, these particles move along 
curved paths, once they come close to pancake configurations
they curve their trajectories, moving almost parallel to them, 
trying to reach the position of filaments.
Again they cannot cross it, so they modify their motion, 
asymptotically approaching them, to finally fall, for $\tau_a \to \infty$, into
the knots corresponding to the minima of the initial 
gravitational potential. 
Altogether, this type of dynamics implies an artificial thickening
of particles around pancakes, filaments and knots, which mimics the 
real gravitational clustering around this type of structures. 
The physical thickening of the particle density around pancakes, filaments
and knots, caused by the damped oscillations around these structures is 
replaced by an approximately exponential slowing down of particle motions, 
which however overestimates the actual particle deceleration. 
In the specular process of evacuation of initially underdense 
regions, FFA overshoots the actual dynamics. 
Provided one gives up resolving the trajectories of individual particles,
these effects produce a density field which looks roughly similar to the real 
one; in this sense the method should be considered intrinsically Eulerian.

Matarrese et al. (1992) have considered the evolution of structures on 
large scales as described by FFA and compared it with the results of ZEL and 
of a $N$--body simulation, assuming a standard Cold Dark Matter 
model. Compared to the PM results, FFA 
recovers all the main structures in the correct places, even though they look 
thicker and the voids appear more empty and conspicuous.
FFA leads to an excess of sub--structure, 
which is left on the way during the evolution instead of being erased 
by the hierarchical clustering process as in the true dynamics.
The structures obtained by the Zel'dovich approximation, instead, are less 
prominent and more fuzzy, as the particles have diffused away from 
the caustic positions after shell--crossing. 
These results suggest that FFA is able to reproduce,
at the statistical level, the clustering properties of the universe even on
scales reached by the non--linear action of gravity. 

\titleb {2.4}{Adhesion approximation}

In the adhesion approximation (e.g. Gurbatov, Saichev \& Shandarin 1989; 
Kofman, Pogosyan \& Shandarin 1990; Weinberg \& Gunn 1990; 
Williams et al. 1991; Kofman et al. 1992) one modifies the Zel'dovich approach 
by adding an artificial viscosity term to Eq.(9), which is thus replaced by 
$$
{d {\bf u} \over d a} = \nu \nabla^2 {\bf u}.
\eqno(18)
$$
The viscosity is introduced to mimic the actual sticking of particles around 
pancakes, caused by the action of gravity even in a collisionless medium. The 
parameter $\nu$ plays the role of a coefficient of kinematical 
viscosity, which controls the thickness of pancakes. 

The previous equation is the vector generalization of the well--known 
non--linear diffusion or Burgers equation of strong turbulence (e.g. Burgers 
1974). 
One can still 
define a velocity potential through ${\bf u} = \nabla 
\Phi$, which can be determined through the Hopf--Cole substitution 
$\Phi = - 2 \nu \ln U$; the scalar field $U$ satisfies the linear
diffusion or Fokker--Planck equation,
$\partial U / \partial \tau_a = \nu \nabla^2 U$,
with the initial condition $U_0({\bf x}) = \exp[-\Phi_0({\bf x})/2\nu]$. The 
resulting velocity potential reads
$$
\Phi({\bf x},\tau_a) = - 2 \nu \ln \biggl[ {1 \over (4\pi \nu \tau_a)^{3/2}}
\int d^3 q \exp \biggl( - {1 \over 2 \nu} S({\bf x},{\bf q}, \tau_a) \biggr) 
\biggr],
\eqno(19)
$$
where one defines the action 
$S({\bf x},{\bf q}, \tau_a) \equiv \Phi_0({\bf q}) + ({\bf x} - 
{\bf q})^2 /2 \tau_a$, satisfying the Hamilton--Jacobi Eq.(12). 
The corresponding velocity field is easily obtained by differentiation;
the Eulerian positions of the particles are found by direct integration 
of the integral equation
$$
{\bf x}({\bf q}, \tau_a) = {\bf q} + \int_0^{\tau_a} d \tau_a' {\bf u}
[{\bf x}({\bf q},\tau_a'),\tau_a'] ,
\eqno(20)
$$
while the density field can be obtained from Eq.(7). 

The Burgers equation is usually considered in the limit of small (but 
non--vanishing) $\nu$, which corresponds to the limit of large 
Reynolds numbers,
${\cal R}_0 = u_0 \ell_0 /\nu$, $u_0$ and $\ell_0$ being the 
characteristic amplitude and scale of the initial velocity field. 
The product $u_0\ell_0$ can be 
estimated either from the {\it rms} initial velocity potential 
smoothed on some scale $R$, $\langle \Phi_0^2(R) \rangle^{1/2}$, if this is 
convergent, or from the square root of the structural function of 
$\Phi_0$
$$
D(r) = \langle [\Phi_0({\bf x}) - \Phi_0({\bf x} + {\bf r})]^2 \rangle =
(1 / \pi^2) \int_0^\infty d k k^2 {\cal P}_\varphi(k) W^2(kR)
[1 - j_0(kr)] ,
\eqno(21)
$$
evaluated at a suitable lag, e.g. $r \approx R$. Here
${\cal P}_\varphi$ is the 
power--spectrum of the initial gravitational potential and $W(kR)$ a suitable 
low--pass filter. 

In the small $\nu$ case the solution takes a simplified form which can be 
obtained from Eq.(11) through a saddle--point approximation,
$$
\Phi({\bf x},\tau_a) \approx - 2 \nu \ln \biggl[ \sum_\alpha 
{\cal J}({\bf q}_\alpha)^{-1/2}
\exp \biggl( - {1 \over 2 \nu} S({\bf x},{\bf q}_\alpha, \tau_a) \biggr) 
\biggr], 
\eqno(22)
$$
where ${\cal J}({\bf q})= |\!| {\bf 1} + \tau_a 
{\bf D}_0({\bf q}) |\!|$, $~{\bf D}_0$ is the deformation tensor and 
${\bf q}_\alpha$ are the Lagrangian points which minimize the action $S$
at given ${\bf x}$ and $\tau_a$. The Zel'dovich approximation is 
recovered in the limit $\nu \to 0$. 
This model has been applied to perform 
numerical simulations of the large--scale structure of the universe 
or to obtain some physical insight into the 
structure formation process in simplified cases. The model allows to 
obtain the skeleton of the large--scale 
matter distribution by a geometrical technique based on the insertion 
of osculating paraboloids into the hypersurface $\varphi_0({\bf q})$. 

Let us finally mention that a detailed statistical comparison of the various 
approximation schemes discussed here against full numerical simulations 
has been recently performed by Sathyaprakash et al. (1995). 

\titlea{3}{Relativistic dynamics of a self--gravitating collisionless fluid}

\titleb {3.1}{General formalism}

We start by writing the Einstein's equation for a perfect fluid of 
irrotational dust. The formalism outlined in this section is discussed 
in greater detail in (Matarrese \& Terranova 1995). 
With the purpose of studying gravitational instability in a FRW background, 
it is convenient to factor out the homogeneous and isotropic FRW expansion 
of the universe and perform a conformal rescaling of the metric with 
conformal factor $a(t)$, the scale--factor of
FRW models, and adopt the conformal time $\tau$, 
defined by $d\tau = dt / a(t)$ ($\tau$, not to be confused with the variable 
$\tau_a$ of the previous section, is proportional to $t^{1/3}$ in the 
Einstein--de Sitter case).

The line--element is then written in the form 
$$
ds^2 = a^2(\tau)\big[ - c^2 d\tau^2 + \gamma_{\alpha\beta}({\bf q}, \tau) 
dq^\alpha d q^\beta \big] \;. 
\eqno(23)
$$
For later convenience let us fix the Lagrangian coordinates $q^\alpha$ to have 
physical dimension of length and the conformal time variable $\tau$ to have 
dimension of time. As a consequence the spatial metric $\gamma_{\alpha\beta}$
is dimensionless, as is the scale--factor $a(\tau)$ which must be 
determined by solving the Friedmann equations for a perfect fluid of dust 
$$
\biggl({a' \over a}\biggr)^2 = {8\pi G \over 3} \varrho_b a^2 - \kappa c^2 \;,
\eqno(24)
$$
$$
2 {a'' \over a} - \biggl({a' \over a}\biggr)^2 + \kappa c^2 = 0 \;,
\eqno(25)
$$
where $\varrho_b(\tau)$ is the background mean density. 
Here primes denote differentiation with respect to the conformal time $\tau$
and $\kappa$ represents the curvature parameter of FRW models, 
which, because of our choice of dimensions, cannot be normalized as usual. 
So, for an Einstein--de Sitter universe $\kappa=0$, but for a closed
(open) model one simply has $\kappa>0$ ($\kappa<0$). 
Let us also note that the curvature parameter is related to a 
Newtonian squared time--scale $\kappa_N$ through $\kappa_N \equiv \kappa c^2$
(e.g. Coles \& Lucchin 1995); in other words $\kappa$ is an 
intrinsically post--Newtonian quantity. 

By subtracting the isotropic Hubble--flow, one introduces a peculiar 
velocity--gradient tensor 
$$
\vartheta^\alpha_{~\beta} = {1 \over 2} \gamma^{\alpha\gamma} 
{\gamma_{\gamma\beta}}' \;.
\eqno(26)
$$

Thanks to the introduction of this tensor one can write the Einstein's 
equations in a more cosmologically convenient form. 
The energy constraint, i.e. the time--time component of the Einstein's 
equations, reads 
$$
\vartheta^2 - \vartheta^\mu_{~\nu} \vartheta^\nu_{~\mu} + 4 {a' \over a} 
\vartheta + c^2 \bigl( {\cal R} - 6 \kappa \bigr) = 16 \pi G a^2 
\varrho_b \delta \;,
\eqno(27)
$$
where ${\cal R}^\alpha_{~\beta}(\gamma)$ 
is the conformal Ricci curvature of the three--space, i.e. that corresponding 
to the metric $\gamma_{\alpha\beta}$; for the background FRW solution 
$\gamma^{FRW}_{\alpha\beta} = (1 + {\kappa\over 4} q^2)^{-2} 
\delta_{\alpha\beta}$, one has ${\cal R}^\alpha_{~\beta}(\gamma^{FRW}) 
= 2 \kappa 
\delta^\alpha_{~\beta}$. We also introduced the density contrast 
$\delta \equiv (\varrho - \varrho_b) /\varrho_b$. 

The momentum constraint, i.e. the time--space components of the Einstein's 
equations, reads 
$$
\vartheta^\alpha_{~\beta||\alpha} = \vartheta_{,\beta} \;. 
\eqno(28)
$$
To avoid excessive proliferation of symbols, the double vertical bars are used 
here and in the following for covariant derivatives in the 
three--space with metric $\gamma_{\alpha\beta}$. 

Finally, after replacing the density from the energy constraint and
subtracting the background contribution, the evolution 
equation, coming from the space--space components of the Einstein's 
equations, becomes 
$$
{\vartheta^\alpha_{~\beta}}' + 2 {a' \over a} \vartheta^\alpha_{~\beta} + 
\vartheta \vartheta^\alpha_{~\beta} + {1 \over 4} 
\biggl( \vartheta^\mu_{~\nu} \vartheta^\nu_{~\mu} - \vartheta^2 \biggr) 
\delta^\alpha_{~\beta} + {c^2 \over 4} \biggl[ 4 {\cal R}^\alpha_{~\beta} 
- \bigl( {\cal R} + 2 \kappa \bigr) \delta^\alpha_{~\beta} \biggr]
= 0 \;. 
\eqno(29)
$$

The Raychaudhuri equation for the evolution of the 
peculiar volume expansion scalar $\vartheta$ becomes 
$$
\vartheta' + {a' \over a} \vartheta + \vartheta^\mu_{~\nu} \vartheta^\nu_{~\mu} 
+ 4 \pi G a^2 \varrho_b \delta =0 \;. 
\eqno(30)
$$
The main advantage of this formalism is that there is only one dimensionless 
(tensor) variable in the equations, namely the spatial metric tensor 
$\gamma_{\alpha\beta}$, which is present with its partial time derivatives 
through $\vartheta^\alpha_{~\beta}$, 
and with its spatial gradients through the spatial Ricci 
curvature ${\cal R}^\alpha_{~\beta}$. The only remaining variable is the 
density contrast which can be written in the form
$$
\delta({\bf q}, \tau) = (1 + \delta_0({\bf q})) \bigl[\gamma({\bf q}, \tau)/ 
\gamma_0 ({\bf q}) \bigr]^{-1/2} - 1 \;,
\eqno(31)
$$
where $\gamma \equiv {\rm det} ~\gamma_{\alpha\beta}$. 
A relevant advantage of having a single tensorial variable, for our purposes, 
is that there can be no extra powers of $c$ hidden in the definition of 
different quantities. 

Our intuitive notion of Eulerian coordinates, involving a universal absolute 
time and globally flat spatial coordinates is intimately Newtonian; 
nevertheless it is possible to construct a local coordinates system 
which reproduces this picture for a suitable set of observers. 
Local Eulerian -- FRW comoving -- coordinates $x^A$ can be introduced, 
related to the Lagrangian ones $q^\alpha$ via the Jacobian matrix 
with elements
$$
{\cal J}^A_{~~\alpha} ({\bf q},\tau) \equiv 
{\partial x^A \over \partial q^\alpha} \equiv 
\delta^A_{~\alpha} + {\cal D}^A_{~\alpha} ({\bf q},\tau) \;, 
\ \ \ \ \ \ A=1,2,3 \;,
\eqno(32)
$$
where ${\cal D}^A_{~\alpha} ({\bf q},\tau)$ is called
deformation tensor. Each matrix element ${\cal J}^A_{~~\alpha}$
labelled by the Eulerian index $A$ can 
be thought as a three--vector, namely a triad, defined on the 
hypersurfaces of constant conformal time. 
They evolve according to 
$$
{{\cal J}^A_{~~\alpha}}' = \vartheta^\gamma_{~\alpha} {\cal J}^A_{~~\gamma} \;,
\eqno(33)
$$
which also follows from the condition of parallel transport of the triads 
relative to ${\bf q}$ along the world--line of the corresponding 
fluid element $D(a {{\cal J}^A_{~~\alpha}}) / D t =0$. 

Our local Eulerian coordinates are such that the 
spatial metric takes the Euclidean form $\delta_{AB}$,
i.e. 
$$
\gamma_{\alpha\beta} ({\bf q},\tau) = \delta_{AB} {\cal J}^A_{~~\alpha} 
({\bf q},\tau) {\cal J}^B_{~~\beta} ({\bf q},\tau) \;. 
\eqno(34)
$$
Correspondingly the matter density can be rewritten in the suggestive form
$$
\varrho ({\bf q},\tau) = \varrho_b(\tau) \bigl( 1 + \delta_0({\bf q}) \bigr)
\bigl[ {\cal J}({\bf q},\tau) / {\cal J}_0 ({\bf q}) \bigr]^{-1} \;,
\eqno(35)
$$
where ${\cal J} \equiv {\rm det} {\cal J}^A_{~~\alpha}$. Note that, contrary to
the Newtonian case, it is generally impossible in GR to fix ${\cal J}_0=1$,
as this would imply that the initial Lagrangian space is conformally flat, 
which is only possible if the initial perturbations vanish. 

\titleb {3.2}{Linear approximation in Lagrangian coordinates}

We are now ready to deal with the linearization of the equations 
obtained above. 
Let us then write the spatial metric tensor of the physical 
(i.e. perturbed) space--time in the form 
$$
\gamma_{\alpha\beta} = {\bar \gamma}_{\alpha\beta} + w_{\alpha\beta} \;,
\eqno(36)
$$
with ${\bar \gamma}_{\alpha\beta}$ the spatial metric of the 
background space -- in our case the maximally symmetric FRW one,
${\bar \gamma}_{\alpha\beta} = \gamma^{FRW}_{\alpha\beta}$ -- and 
$w_{\alpha\beta}$ a small perturbation. The only
non--geometric quantity in our equations, namely the initial density contrast
$\delta_0$, can be assumed to be much smaller than unity. 

As usual, one can take advantage of the maximal symmetry of the background FRW 
spatial sections to classify metric perturbations as scalars, vectors and 
tensors. One then writes
$$
w_{\alpha\beta} = \chi {\bar \gamma}_{\alpha\beta} + \zeta_{|\alpha\beta} +
{1 \over 2} \bigl( \xi_{\alpha | \beta} + \xi_{\beta | \alpha} \bigr) 
+ \pi_{\alpha\beta} \;, 
\eqno(37)
$$
with
$$
\xi^\alpha_{~~|\alpha}= \pi^\alpha_{~\alpha} = \pi^\alpha_{~\beta|\alpha}=0 \;,
\eqno(38)
$$
where a single vertical bar is used for covariant differentiation in the 
background three--space with metric ${\bar \gamma}_{\alpha\beta}$. 
In the above decomposition $\chi$ and $\zeta$ represent scalar modes, 
$\xi^\alpha$ vector modes and $\pi^\alpha_{~\beta}$ tensor modes 
(indices being raised by the contravariant background three--metric). 

Before entering into the discussion of the equations for these perturbation 
modes, let us quote a result which will be also useful later. 
In the $\vartheta^\alpha_{~\beta}$ evolution equation and in the energy 
constraint the combination ${\cal P}^\alpha_{~\beta} \equiv 
4 {\cal R}^\alpha_{~\beta} - \bigl( {\cal R} + 
2 \kappa \bigr) \delta^\alpha_{~\beta}$ and its trace appear. 
To first order in the metric perturbation one has
$$
{\cal P}^\alpha_{~\beta} (w) = -2 \biggl[ \bigl(\nabla^2 - 2 \kappa \bigr) 
\pi^\alpha_{~\beta} + {\chi_|}^\alpha_{~\beta} + \kappa \chi 
\delta^\alpha_{~\beta} \biggr] \;,
\eqno(39)
$$
where $\nabla^2 ( \cdot ) \equiv {( \cdot )_|}^\gamma_{~\gamma}$. 
Only the scalar mode $\chi$ and the tensor modes contribute to the 
three--dimensional Ricci curvature. 

As well known, in linear theory scalar, vector and tensor modes are 
independent. The equation of motion for the tensor modes 
is obtained by linearizing the traceless part of the 
$\vartheta^\alpha_{~\beta}$ evolution equation. One has
$$
{\pi_{\alpha\beta}}'' + 2 {a' \over a} {\pi_{\alpha\beta}}' 
- c^2 \bigl(\nabla^2 - 2 \kappa \bigr) \pi_{\alpha\beta} = 0 \;,
\eqno(40)
$$
which is the equation for the free propagation of gravitational 
waves in a FRW background. The general solution of this equation is 
well--known (e.g. Weinberg 1972) and will not be reported here. 

At the linear level, in the irrotational case, the two vector modes represent 
gauge modes which can be set to zero, $\xi^\alpha=0$. 

The two scalar modes are linked together through the momentum constraint, 
which leads to the relation
$\chi = \chi_0 + \kappa (\zeta - \zeta_0)$.
The energy constraint gives
$$
\bigl( \nabla^2 + 3 \kappa \bigr) \biggl[ {a' \over a} \zeta' + 
\bigl(4 \pi G a^2 \varrho_b - \kappa c^2 \bigr) 
(\zeta - \zeta_0) - c^2 \chi_0 \biggr] = 8 \pi G a^2 \varrho_b \delta_0 \;, 
\eqno(41)
$$
while the evolution equation gives
$$
\zeta'' + 2 {a' \over a} \zeta' = c^2 \chi \;.
\eqno(42)
$$

An evolution equation only for the scalar mode $\zeta$ can be obtained 
by combining together the evolution equation and the energy constraint;
it reads 
$$
\bigl( \nabla^2 + 3 \kappa \bigr) \biggl[ \zeta'' + {a' \over a} \zeta' 
- 4\pi G a^2 \varrho_b (\zeta - \zeta_0) \biggr] = - 8 \pi G a^2 
\varrho_b \delta_0 \;. 
\eqno(43)
$$

On the other hand, linearizing the solution of the continuity 
equation, gives 
$$
\delta = \delta_0 - {1 \over 2} (\nabla^2 + 3 \kappa ) (\zeta - \zeta_0) \;,
\eqno(44)
$$
which replaced in the previous equation gives 
$$
\delta'' + {a' \over a} \delta' - 4\pi G a^2 \varrho_b \delta = 0 \;.
\eqno(45)
$$
This is the well--known equation for linear density fluctuation, whose 
general solution can be found in (Peebles 1980). 
Once $\delta(\tau)$ is known, 
one can easily obtain $\zeta$ and $\chi$, which completely solves the linear 
problem. 

Eq.(43) above has been obtained in whole generality; one could have used 
instead the well--known residual gauge ambiguity of the synchronous 
coordinates to simplify its form. In fact, $\zeta$ is determined up to a 
space--dependent scalar, which would neither contribute to the spatial 
curvature, nor to the velocity--gradient tensor. For instance, one
could fix $\zeta_0$ so that $(\nabla^2 + 3 \kappa) \zeta_0 = - 2 \delta_0$,
so that the $\zeta$ evolution equation takes the same form as that for 
$\delta$. 

In order to better understand the physical meaning of the two 
scalar modes $\chi$ and $\zeta$, let us consider the simplest case of an 
Einstein--de Sitter background ($\kappa=0$), for which 
$a(\tau) \propto \tau^2$. By fixing the gauge so that 
$\nabla^2 \zeta_0 = - 2 \delta_0$ one obtains $\chi(\tau)=\chi_0$ and 
$$
\zeta(\tau) = {c^2 \over 10} \chi_0 \tau^2 + B_0 \tau^{-3} \;,
\eqno(46)
$$
where the amplitude $B_0$ of the decaying mode is an arbitrary function of the 
spatial coordinates. Consistency with the Newtonian limit suggests 
$\chi_0 \equiv - {10 \over 3c^2} \varphi_0$, with $\varphi_0$ the initial 
peculiar gravitational potential, related to $\delta_0$ through 
$\nabla^2 \varphi_0 = 4 \pi G a_0^2 \varrho_{0b} \delta_0$. One can then write 
$$
\zeta(\tau) = - {1 \over 3} \varphi_0 \tau^2 + B_0 \tau^{-3} \;.
\eqno(47)
$$

This result clearly shows that, at the Newtonian level, the linearized 
metric is $\gamma_{\alpha\beta} = \delta_{\alpha\beta} + \zeta_{|\alpha\beta}$,
while the perturbation mode $\chi$ is already 
post--Newtonian. 

These results also confirm the above conclusion that in the general GR case 
the initial Lagrangian spatial metric cannot be flat, i.e. 
${\cal J}_0 \neq 1$, because of the initial ``seed" post--Newtonian metric 
perturbation $\chi_0$. 

\titleb {3.3}{Recovering the Newtonian approximation in the Lagrangian picture} 

The Newtonian equations in Lagrangian form can be obtained from the 
full GR equations by an expansion in inverse 
powers of the speed of light; as a consequence of our
gauge choice, however, no odd powers of $c$ appear in the 
equations, which implies that the expansion parameter can be 
taken to be $1/c^2$. 

Let us then expand the spatial metric in a form analogous to that 
used in the linear perturbation analysis above. 
$$
\gamma_{\alpha\beta} = {\bar \gamma}_{\alpha\beta} + {\cal O} 
\biggl( {1 \over c^2} \biggr) \;. 
\eqno(48)
$$

To lowest order in our expansion, the evolution equation
and the energy constraint imply that 
${\bar {\cal P}}^\alpha_{~\beta} \equiv 
{\cal P}^\alpha_{~\beta} ({\bar \gamma}) =0$, and recalling that 
$\kappa=\kappa_N/c^2$, one gets 
${\bar {\cal R}}^\alpha_{~\beta} \equiv {\cal R}^\alpha_{~\beta} 
({\bar \gamma}) = 0$: 
in the Newtonian limit the spatial curvature identically vanishes. 
This important conclusion implies that 
${\bar \gamma}_{\alpha\beta}$ can be transformed to $\delta_{AB}$
globally, i.e. that one can write 
${\bar \gamma}_{\alpha\beta} = \delta_{AB} {\bar {\cal J}}^A_{~~\alpha} 
{\bar {\cal J}}^B_{~~\beta}$,
with integrable Jacobian matrix coefficients. In other words, at each time 
$\tau$ there exist global Eulerian coordinates $x^A$ such that 
$$
{\bf x}({\bf q}, \tau) = {\bf q} + {\bf S}({\bf q}, \tau) \;,
\eqno(49)
$$
where ${\bf S}({\bf q}, \tau)$ is called the displacement vector, 
and the deformation tensor becomes in this limit 
$$
{\bar {\cal D}}^A_{~\alpha} = {\partial S^A \over \partial q^\alpha} \;.
\eqno(50)
$$

The Newtonian Lagrangian metric can therefore be written in the form 
$$
{\bar \gamma}_{\alpha\beta}({\bf q}, \tau) = \delta_{AB} 
\biggl(\delta^A_{~\alpha} + {\partial S^A({\bf q}, \tau) 
 \over \partial q^\alpha} \biggr) 
\biggl(\delta^B_{~\beta} + {\partial S^B({\bf q}, \tau) 
\over \partial q^\beta} \biggr) \;.
\eqno(51)
$$

One can rephrase the above result as follows: the Lagrangian spatial metric in 
the Newtonian limit is that of Euclidean three--space in time--dependent 
curvilinear coordinates $q^\alpha$, defined at each time $\tau$ in terms of 
the Eulerian ones $x^A$ by inversion. As a consequence, the 
Christoffel symbols involved in spatial covariant derivatives (which will be
indicated by a single bar or by a nabla operator followed by greek indices) 
do not vanish, but the vanishing of the spatial curvature implies that 
these covariant derivatives always commute. 

Contrary to the evolution equation and the energy constraint, the 
Raychaudhuri equation and the momentum constraint 
contain no explicit powers of $c$, and therefore preserve their form in going 
to the Newtonian limit. These equations therefore determine the background 
Newtonian metric ${\bar \gamma}_{\alpha\beta}$, i.e. they govern the evolution 
of the displacement vector ${\bf S}$. 

The Raychaudhuri equation becomes the master equation for the Newtonian 
evolution; it takes the form 
$$
{\bar \vartheta}' + {a' \over a} {\bar \vartheta} + 
{\bar \vartheta}^\mu_{~\nu} {\bar \vartheta}^\nu_{~\mu} + 
4 \pi G a^2 \varrho_b \bigl( {\bar \gamma}^{-1/2} - 1 \bigr) = 0 \;,
\eqno(52)
$$
where
$$
{\bar \vartheta}^\alpha_{~\beta} \equiv {1 \over 2} 
{\bar \gamma}^{\alpha\gamma} {\bar \gamma_{\gamma\beta}}' \;, 
\eqno(53)
$$
and, for simplicity, $\delta_0=0$ was assumed (a restriction which is, however,
not at all mandatory). We also used the residual gauge freedom 
of our coordinate system to set ${\bar \gamma}_{\alpha\beta}(\tau_0) = 
\delta_{\alpha\beta}$, implying ${\bar {\cal J}}_0=1$, i.e. to make 
Lagrangian and Eulerian coordinates coincide at the initial time. 
That this choice is indeed possible in the Newtonian limit can be understood 
from our previous linear analysis, where this is achieved by taking, e.g., 
$\zeta_0=0$. 

The momentum constraint, ${\bar \vartheta}^\mu_{~\nu |\mu } = 
{\bar \vartheta}_{,\nu}$, 
is actually related to the irrotationality assumption. 

Let us also notice a general property of our 
expression for the Lagrangian metric: at each time $\tau$ it can be 
diagonalized by going to the local and instantaneous principal axes
of the deformation tensor. Calling $\bar \gamma_\alpha$ 
the eigenvalues of the 
metric tensor, ${\bar {\cal J}}_\alpha$ those of the Jacobian and 
${\bar d}_\alpha$ those of the deformation tensor, 
one has
$$
\bar \gamma_\alpha ({\bf q},\tau) = {\bar {\cal J}}_\alpha^2({\bf q},\tau) = 
\bigl( 1 + {\bar d}_\alpha ({\bf q},\tau) \bigr)^2 \;.
\eqno(54)
$$

From this expression it becomes evident that, at shell--crossing, 
where some of the Jacobian eigenvalues go to zero, the related covariant 
metric eigenvalues just vanish. On the other hand, other quantities, like 
the matter density, the peculiar volume expansion scalar and some eigenvalues 
of the shear and tidal tensor will generally diverge at the location of the 
caustics. 
This diverging behaviour makes the description of the system 
extremely involved after this event. Although dealing with this problem is 
far outside the aim of the present notes. let us just mention that a number of 
ways out are available. One can convolve the various dynamical variables by a 
suitable low--pass filter, either at the initial time, in order to postpone 
the occurrence of shell--crossing singularities, or at the time when they 
form, in order to smooth the singular behaviour; alternatively one 
can abandon the perfect fluid picture and resort to a discrete point--like 
particle set, which automatically eliminates the possible occurrence of 
caustics, at least for generic initial data. 
At this level, anyway, we prefer to take a conservative 
point of view and assume that the  actual range of validity of this formalism 
is up to shell--crossing. 

\titleb {3.4}{Recovering the Newtonian approximation in the Eulerian picture} 

As demonstrated above, it is always 
possible, in the frame of the Newtonian approximation, to define a global 
Eulerian picture. This will be the picture of the fluid evolution as given 
by an observer that, at the point ${\bf x} = {\bf q} + {\bf S}({\bf q}, \tau)$ 
and at the time $\tau$ observes the fluid moving with physical peculiar 
three--velocity ${\bf v} = d {\bf S}/d \tau$. From the point
of view of a Lagrangian observer, who is comoving with the fluid, 
the Eulerian observer, which is located at constant ${\bf x}$, is moving with 
three--velocity $d {\bf q} ({\bf x}, \tau)/d \tau = - {\bf v}$. 

The line--element characterizing the Newtonian approximation in the 
Eulerian frame is well--known (e.g. Peebles 1980)
$$
ds^2 = a^2(\tau) \biggl[ - \biggl(1 + {2\varphi_g ({\bf x}, \tau) \over c^2} 
\biggr) ~c^2 d\tau^2 + \delta_{AB} dx^A dx^B \biggr] \;, 
\eqno(55)
$$
with $\varphi_g$ the peculiar gravitational potential, determined 
by the mass distribution through the Eulerian Poisson equation, 
$$
\nabla_x^2 \varphi_g ({\bf x}, \tau) = 4 
\pi G a^2(\tau) \varrho_b(\tau) \delta({\bf x}, \tau) \;,
\eqno(56)
$$
where the Laplacian $\nabla_x^2$, as well as the nabla operator $\nabla$, 
have their standard Euclidean meaning. 
The perturbation in the time--time component of the metric tensor 
here comes from the different proper time of the Eulerian 
and Lagrangian observers. 

It is now crucial to realize that all the dynamical equations obtained 
so far, being entirely expressed in terms of three--tensors, keep their form 
in going to the Eulerian picture, only provided the convective time 
derivatives of tensors of any rank (scalars, vectors and 
tensors) are modified as follows: 
$$
{D \over D \tau} \rightarrow {\partial \over \partial \tau} + 
{\bf v} \cdot \nabla \;, \ \ \ \ \ \ \ 
{\bf v} \equiv {d {\bf S} \over d \tau} \;.
\eqno(57)
$$

This follows from the fact that, for the metric above, $\bar \Gamma^0_{AB} = 
\bar \Gamma^A_{0B} =\bar \Gamma^A_{BC}=0$, which also obviously implies that 
covariant derivatives with respect to $x^A$ reduce to partial ones. 

The irrotationality assumption now has the obvious consequence that we can 
define an Eulerian velocity potential $\Phi_v$ through
$$
{\bf v} ({\bf x}, \tau) = \nabla \Phi_v ({\bf x}, \tau) \;.
\eqno(58)
$$
The Newtonian peculiar velocity--gradient tensor then becomes 
$$
\bar \vartheta_{AB} = {\partial^2 \Phi_v \over \partial x^A \partial x^B} \;,
\eqno(59)
$$
because of which the momentum constraint gets trivially satisfied and the 
magnetic Weyl tensor becomes identically zero in the Newtonian limit. 

We can now write the Raychaudhuri equation for the Eulerian peculiar 
volume expansion scalar $\bar \vartheta$, 
and use the Poisson equation to get, as a first spatial integral, the 
Euler equation
$$
{\bf v}~' + {\bf v} \cdot \nabla {\bf v} + 
{a' \over a} {\bf v} = - \nabla \varphi_g \;.
\eqno(60)
$$

This can be further integrated to give the Bernoulli equation
$$
\Phi_v' + {a' \over a} \Phi_v + {1\over 2} \bigl(\nabla \Phi_v \bigr)^2 
= - \varphi_g \;.
\eqno(61)
$$

All these equations would of course recover the form of Section 2, 
if the time variable and the peculiar velocity were rescaled as described at 
the beginning of that section. 
Having shown the equivalence of this method, in the Newtonian limit, with 
the standard one, it would be also trivial to recover the Zel'dovich 
approximation in this frame. This point is further discussed by Matarrese
\& Terranova (1995). 

\titlea{4}{Conclusions and discussion}

Aim of these notes was to introduce the reader to the 
theory of the dynamics of cosmological perturbations beyond the linear 
approximation. Let me spend this final section to discuss 
what I consider an open issue in this field. The issue is whether there exists 
a range of scales where relativistic effects and non--linear 
evolution both come into play. 
The standard Newtonian paradigm states that the lowest scale at which the 
approximation can be reasonably applied is set by the amplitude of the 
gravitational potential and is given by the Schwarzschild radius of the 
collapsing body, which is negligibly small for any relevant cosmological mass 
scale. 
What is completely missing in this criterion is the role of the shear, which 
causes the presence of non--scalar contributions to the metric perturbations. 
A non--vanishing shear component is, in fact, an unavoidable feature of 
realistic cosmological perturbations and affects the dynamics 
in at least three ways, all related to non--local effects, i.e. to the 
interaction of a given fluid element with the environment. 
First, at the lowest perturbative order the shear is related to the 
tidal field generated by the surrounding material by a simple proportionality 
law. Second, it is related to a dynamical tidal induction: the modification 
of the environment forces the fluid element to modify its shape and density. 
Third, and most important here, a non--vanishing shear field leads to the 
generation of a traceless and divergenceless metric perturbation which can be 
understood as gravitational radiation emitted by non--linear perturbations. 
Note that the two latter effects are only detected if one 
allows for non--scalar perturbations in physical quantities. 
Truly tensor perturbations are in fact dynamically generated 
by the gravitational instability of initially scalar perturbations, 
independently of the initial presence of gravitational waves. 

Using a post--Newtonian expansion in Lagrangian coordinates, 
Matarrese \& Terranova (1995) obtained a general formula 
for the tensor modes $\pi_{AB}$ produced by non--linear evolving 
perturbations. In the standard case, where the cosmological perturbations form 
a homogeneous and isotropic random field, they obtained a heuristic 
perturbative estimate of their amplitude in terms of the 
{\it rms} density contrast and of the ratio of the typical perturbation scale 
$\lambda$ to the Hubble radius $r_H=c H^{-1}$ (where $H$ is Hubble's 
constant). They found $\pi_{rms} / c^2 \sim 
\delta_{rms}^2 (\lambda / r_H )^2$. 
These tensor modes give rise to a stochastic background of 
gravitational waves which gets a non--negligible amplitude in 
the so--called {\it extremely--low--frequency} band (e.g. Thorne 1995), 
around $10^{-14}$ --  $10^{-15}$ Hz. 
One can roughly estimate that the present--day closure density of this
gravitational--wave background would be $\Omega_{gw}(\lambda) \sim 
\delta_{rms}^4 (\lambda / r_H)^2$. 
In standard scenarios for the formation of structure in the universe, 
the closure density on scales 
$1  - 10$ Mpc would be $\Omega_{gw} \sim 10^{-5} - 10^{-6}$. 
The amplitude of this gravitational--wave contribution, 
$\pi \sim \delta^2 (\lambda /r_H)^2$, is an important counter--example to the 
standard paradigm stated above, according to which relativistic effects 
should be proportional to $\varphi_g/c^2 \sim \delta (\lambda / r_H)^2$.

\bigskip\noindent

\begrefchapter{References}
\ref Buchert, T. 1995, to appear in Proc. Enrico Fermi School, 
Course CXXXII, {\it Dark Matter in the Universe}, Varenna 1995, preprint 
astro-ph/9509005. 
\ref Burgers, J.M. 1974, {\it The Nonlinear Diffusion Equation},
Dordrecht: Reidel. 
\ref Catelan, P., Lucchin, F., Matarrese, S. \& Moscardini, L. 
1995, MNRAS, 276, 39. 
\ref Coles, P. \& Lucchin, F. 1995, {\it Cosmology: The Origin and Evolution of
Cosmic Structure}, Chichester: Wiley. 
\ref Efstathiou, G., Davis, M., Frenk, C. \& White S.D.M. 1985,  
ApJS, 57, 241. 
\ref Gurbatov, S.N., Saichev, A.I. \& Shandarin, S.F. 1989, MNRAS, 236, 385. 
ref Hockney, R.W. \& Eastwood, J.W. 1981, {\it Computer Simulations
using Particles}, New York: McGraw--Hill. 
\ref Kofman, L., Pogosyan, D. \& Shandarin, S. 1990, MNRAS 242, 200. 
\ref Kofman, L., Pogosyan, D., Melott, A. \& Shandarin, S. 1992, ApJ, 393, 437. 
\ref Matarrese, S., Lucchin, F., Moscardini, L., Saez, D. 1992, MNRAS, 259, 437.
\ref Matarrese, S. \& Terranova, D. 1995, preprint astro-ph/9511093. 
\ref Nusser, A., Dekel, A., Bertschinger, E. \& Blumenthal, G.R. 1991, ApJ, 
379, 6. 
\ref Peebles, P.J.E. 1980, {\it The Large--Scale Structure of the Universe}, 
Princeton: Princeton University Press. 
\ref Sathyaprakash, B.S., Sahni, V., Munshi, D., Pogosyan, D. \& Melott, A.L. 
1995, MNRAS, 275, 463. 
\ref Thorne, K.S. 1995, in Proc. {\it Snowmass 95 Summer Study on 
Particle and Nuclear Astrophysics and Cosmology}, edited by Kolb E.W. \& 
Peccei R., preprint gr-qc/9506086. 
\ref Weinberg, S. 1972, {\it Gravitation and Cosmology}, New York: Wiley. 
\ref Weinberg, D.H. \& Gunn, J.E. 1990, MNRAS, 247, 260. 
ref Williams, B.G., Heavens, A.F., Peacock, J.A. \& Shandarin, S.F. 1991, 
MNRAS, 250, 458. 
\ref Zel'dovich, Ya.B. 1970, A\&A, 5, 84. 
\endref
\byebye